\RequirePackage{fixltx2e}
\documentclass[pre,reprint,amssymb,amsmath,amsfonts,aps,floatfix,superscriptaddress, a4paper]{revtex4-1}
\usepackage{graphicx, bm, multirow, rotating, units, upgreek, xspace, ucs, enumerate, hyperref}
\hypersetup{colorlinks,breaklinks, citecolor=[rgb]{0,0.5,0}, linkcolor=[rgb]{0.5,0,0}, urlcolor=[rgb]{0.5,0,0.5}}

\newcommand {\bR}{{\mathbf{R}}}
\newcommand {\bA}{{\mathbf{A}}}
\newcommand {\bL}{{\bm{\mathcal{L}}}}

\begin{document}

\title{Evolving functional network properties and synchronizability during human epileptic seizures}
\author{Kaspar A. \surname{Schindler}}%
\affiliation{Department of Epileptology, University of Bonn, Sigmund-Freud-Str. 25, 53105 Bonn, Germany} 
\affiliation{Department of Neurology, Inselspital, Bern University Hospital, and University of Bern, Switzerland}

\author{Stephan \surname{Bialonski}}
\affiliation{Department of Epileptology, University of Bonn, Sigmund-Freud-Str. 25, 53105 Bonn, Germany}
\affiliation{Helmholtz-Institute for Radiation and Nuclear Physics, University of Bonn, Nussallee 14-16, 53115 Bonn, Germany}

\author{Marie-Therese \surname{Horstmann}}
\affiliation{Department of Epileptology, University of Bonn, Sigmund-Freud-Str. 25, 53105 Bonn, Germany}
\affiliation{Helmholtz-Institute for Radiation and Nuclear Physics, University of Bonn, Nussallee 14-16, 53115 Bonn, Germany}
\affiliation{Interdisciplinary Center for Complex Systems, University of Bonn, R{\"o}merstr. 164, 53117 Bonn, Germany}

\author{Christian E. \surname{Elger}}%
\affiliation{Department of Epileptology, University of Bonn, Sigmund-Freud-Str. 25, 53105 Bonn, Germany} 

\author{Klaus \surname{Lehnertz}}
\affiliation{Department of Epileptology, University of Bonn, Sigmund-Freud-Str. 25, 53105 Bonn, Germany}
\affiliation{Helmholtz-Institute for Radiation and Nuclear Physics, University of Bonn, Nussallee 14--16, 53115 Bonn, Germany}
\affiliation{Interdisciplinary Center for Complex Systems, University of Bonn, R{\"o}merstr. 164, 53117 Bonn, Germany} % 

\begin{abstract}
We assess electrical brain dynamics before, during, and after one-hundred human epileptic seizures with different anatomical onset locations by 
statistical and spectral properties of functionally defined networks. We observe a concave-like temporal evolution of characteristic path length and cluster coefficient indicative of a movement from a more random toward a more regular and then back toward a more random functional topology. Surprisingly, synchronizability was significantly decreased during the seizure state but increased already prior to seizure end. Our findings underline the high relevance of studying complex systems from the view point of complex networks, which may help to gain deeper insights into the complicated dynamics underlying epileptic seizures.
\end{abstract}
%\pacs{87.19.lm, 05.45.Xt, 89.75.Hc, 05.45.Tp} 
\maketitle

%lead paragraph
\textbf{Epilepsy represents one of the most common neurological disorders, second only to stroke. Patients live with a considerable risk to sustain serious or even fatal injury during seizures. In order to develop more efficient therapies the pathophysiology underlying epileptic seizures should be better understood.  In human epilepsy, however, the exact mechanisms underlying seizure termination are still as uncertain as are mechanisms underlying seizure initiation and spreading. There is now growing evidence that an improved understanding of seizure dynamics can be achieved when considering epileptic seizures as network phenomena. By applying graph-theoretical concepts, we analyzed seizures on the EEG from a large patient group and observed that a global increase of neuronal synchronization prior to seizure end may be promoted by the underlying functional topology of epileptic brain dynamics. This may be considered as an emergent self-regulatory mechanism for seizure termination, providing clues as to how to efficiently control seizure networks.}

\section{Introduction}
Complex networks can be observed in a wide variety of natural and man-made systems \cite{Watts1998,Albert2002,Newman2003,Milo2004,Boccaletti2006a}, and an important general problem is the relationship between the connection structure and the dynamics of these networks. With graph-theoretical approaches, networks may be characterized using graphs, where nodes represent the elements of a complex system and edges their interactions. In the study of brain dynamics \cite{Bassett2006b,Reijneveld2007}, a node may represent the dynamics of a circumscribed brain region determined by electrophysiologic \cite{Bassett2006,Stam2007a,Ponten2007} or imaging techniques \cite{Eguiluz2005,Salvador2005,Achard2006}. Then two nodes are connected by an edge, or direct path, if the strength of their interaction increases above some threshold. Among other structural (or statistical) parameters, the average shortest path length $L$ and the cluster coefficient $C$ are important characteristics of a graph \cite{Watts1998,Strogatz2001}. $L$ is the average fewest number of steps it takes to get from each node to every other, and is thus an emergent property of a graph indicating how compactly its nodes are interconnected. $C$ is the average probability that any pair of nodes is linked to a third common node by a single edge, and thus describes the tendency of its nodes to form local clusters. High values of both $L$ and $C$ are found in regular graphs, in which neighboring nodes are always interconnected yet it takes many steps to get from one node to the majority of other nodes, which are not close neighbors. At the other extreme, if the nodes are instead interconnected completely at random, both $L$ and $C$ will be low. 

Recently, the emergence of collective dynamics in complex networks has been intensively investigated in various fields \cite{Barahona2002,Nishikawa2003,Timme2004,Hong2004,Roxin2004,Belykh2005a,Motter2005,Donetti2005,Chavez2005,Chavez2006a,Atay2005,Arenas2006,Motter2006,Zhou2006a,Zhou2006,Zhou2006b,Gomez2007}. 
It has for example been proposed that random, small-world, and scale-free networks, due to their small network distances, might support efficient and stable globally synchronized dynamics \cite{Lago2000,Barahona2002,Hong2002}. Synchronized dynamics, however, depends not only on statistical but also on spectral properties of a network, which can be derived from the eigenvalue spectrum of the Laplacian matrix describing the corresponding network \cite{Atay2006}. Although a number of studies reported on a correlation between statistical network properties (such as degree homogeneity, cluster coefficient, and degree distribution) and network synchronizability, the exact relationship between the propensity for synchronization of a network and its topology has not yet been fully clarified.

One of the most challenging dynamical systems in nature is the human brain, a large, interacting, complex network with nontrivial topological properties \cite{Sporns2004,Eguiluz2005,Achard2006,He2007}. Anatomical data, theoretical considerations, and computer simulations suggest that brain networks exhibit high levels of clustering combined with short average path lengths, which was taken as an indication of a small-world architecture \cite{Hilgetag2000,Bassett2006b,Sporns2007}. A disorder of the brain that is known to be particularly associated with changes of neuronal synchronization is epilepsy along with its cardinal symptom, recurrent epileptic seizures.  Seizures are extreme events with transient, strongly enhanced collective activity of spatially extended neuronal networks \cite{McCormick2001,Lehnertz2006}. Despite considerable progress in understanding the physiological processes underlying epileptic dynamics, the network mechanisms involved in the generation, maintenance, propagation, and termination of epileptic seizures in humans are still not fully understood. There are strong indications that seizures resemble a nonlinear deterministic dynamics \cite{Andrzejak2001b}, and recent modeling studies \cite{Buzsaki2004b,Netoff2004,Percha2005,Dyhrfjeld-Johnsen2007,Feldt2007,Morgan2008} indicate the general importance of network topology in epilepsy.  Clinical and anatomic observations together with invasive electroencephalography and functional neuroimaging now provide increasing evidence for the existence of specific cortical and subcortical \emph{epileptic networks} in the genesis and expression of not only primary generalized but also focal onset seizures \cite{Bertram1998,Bragin2000,Bartolomei2001,Spencer2002,Blumenfeld2004,Guye2006,Ponten2007,Gotman2008,Luat2008}. An improved understanding of both structure and dynamics of epileptic networks underlying seizure generation could improve diagnosis and, more importantly, could advice new treatment strategies, particularly for the 25\,\% of patients whose seizures cannot be controlled by any available therapy.

In order to gain deeper insights into the global network dynamics during seizures we study -- in a time resolved manner -- statistical and spectral properties of functionally defined seizure networks in human epileptic brains. We observe that, while seizures evolve, statistical network properties indicate a concave-like movement between a more regular (during seizures) and a more random functional topology (prior to seizure initiation and already before seizure termination). Network synchronizability, however, is drastically decreased during the seizure state and increases already prior to seizure end. We speculate that network randomization, accompanied by an increasing synchronization of neuronal activity may be considered as an emergent self-regulatory mechanism for seizure termination.

\section{Methods}
\label{methods}
We retrospectively analyzed multichannel ($n=53 \pm 21$ channels) electroencephalograms (EEG) that were recorded prior to, during, and after one-hundred focal onset epileptic seizures from 60 patients undergoing pre-surgical evaluation for drug-resistant epilepsy. Seizure onsets were localized in different anatomical regions. All patients had signed informed consent that their clinical data might be used and published for research purposes. The study protocol had previously been approved by the ethics committee of the University of Bonn. EEG data were recorded via chronically implanted strip, grid, or depth electrodes from the cortex and from within relevant structures of the brain, hence with a high signal-to-noise ratio. Signals were sampled at 200 Hz using a 16 bit analog-to-digital (A/D) converter and filtered within a frequency band of 0.5 to 70 Hz. In order to minimize the influence of a particular referencing of electrodes on spatially extended correlations, we here applied a bipolar re-referencing prior to analyses by calculating differences between signals from nearest neighbor channels. 

\begin{figure}
  \includegraphics[width = 3.375 in]{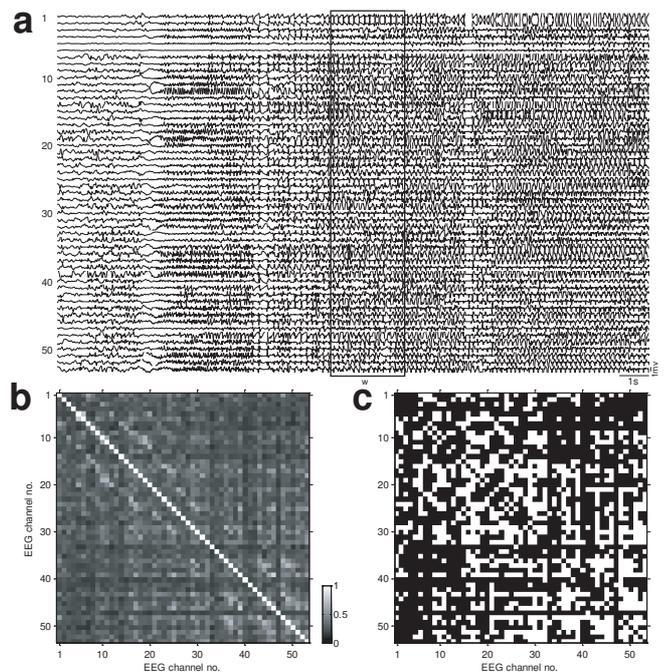}
  \caption{Deriving the adjacency matrix. \textbf{a}, Representative EEG recording with 53 bipolar channels during the initial part of a seizure. \textbf{b}, The maximal lag-correlation matrix $\bR$ is computed for a sliding window $w$ of 2.5s (500 sampling points). The adjacency matrix is constructed by adaptive thresholding in such a way that the resulting binary adjacency matrix $\bA$ (shown in \textbf{c}) represents a connected graph. Each white square indicates that the corresponding nodes (EEG signals) are connected.}
  \label{fig:figure1}
\end{figure} 

We defined \emph{functional} links \cite{Dodel2002,Eguiluz2005,Stam2007a,Ponten2007} between any pair of bipolar EEG channels $i$ and $j$ ($i,j \in \left[1,n\right]$) -- regardless of their anatomical connectivity -- using the cross-correlation function as a simple and most commonly used measure for interdependence between two EEG signals \cite{Brazier1956,Bertashius1991,QuianQuiroga2002,Mormann2003a,Kreuz2007}, which is both computationally efficient and, in light of the known correlation-based changes of synaptic strengths \cite{Cooke2006}, also physiologically plausible. We used a sliding window approach (see Fig. \ref{fig:figure1}) to estimate the elements $\rho^{\rm max}_{ij}$ of the normalized maximum-lag correlation matrix $\bR$

\begin{equation}
\rho^{\rm max}_{ij} = \max \limits_{\tau} \left\{\frac{C(x_i,x_j)(\tau)}{\sqrt{C(x_i,x_i)(0)C(x_j,x_j)(0)}}\right\}
\label{eq:rho}
\end{equation}

where
\begin{equation}
C(x_i,x_j)(\tau)= 
	\left\{
		\begin{array}{ll}
			\sum_{t=1}^{w_l-\tau} x_i(t+\tau)x_j(t), & \tau \geq 0 \\
			C(x_j,x_i)(-\tau), & \tau < 0
		\end{array}
	\right.
\end{equation}

denotes the cross-correlation function. This function yields high values for such time lags $\tau$ for which the signals $x_i(t+\tau)$ and $x_j(t)$ have a similar course in time. The sliding window $w$ had a duration $w_l$ = 2.5 s (500 sampling points; no overlap), which can be regarded as a compromise between the required statistical accuracy for the calculation of the cross-correlation function and approximate stationarity within a window's length. $x_i(t)$ denotes the normalized (zero mean and unit variance) EEG signal at channel $i$. From time-resolved matrices $\bR(w)$ we derived adjacency matrices $\bA(w)$ by thresholding:

\begin{equation}
a_{ij} = \left\{              	
		\begin{array}{r@{\quad\quad}l}                   
			1, & i\neq j \wedge \rho^{\rm max}_{ij} \geq T(w),\\                   
			0, & \mbox{otherwise} 
	  \end{array}       
\right. 
\label{eq:admat}
\end{equation}

Elements on the main diagonal of $\bA$ were set to 0 in order to exclude self-connections of nodes. Since our aim is to characterize the evolving \emph{global} network dynamics, we chose, for each time window, the highest possible threshold for which the resulting graph represented by $\bA(w)$ was connected.  Starting from a threshold $T(w)=1$ we gradually decreased $T(w)$, and we calculated, at each step, the second smallest eigenvalue $\lambda_\text{min}$ of the corresponding Laplacian matrix $\bL$, whose elements are ${\cal L}_{ij} = k_i\delta_{ij} - a_{ij}$, where $\delta_{ij}$ is the Kronecker delta, and $k_i$ denotes the degree of node $i$. $\lambda_\text{min}$ is positive if and only if the graph is connected \cite{Atay2005}. 

Following Ref. \cite{Latora2001} we used $\bA$ to compute average shortest path length $L$, cluster coefficient $C$, and normalized edge density $\epsilon$ (i.e, the actual number of edges in $\bA$ divided by the number of possible edges between $n$ nodes) and assigned their values to the time point at the beginning of each window $w$. To detect deviations from a random network topology we consider the ratios $C/C_\text{r}$ and $L/L_\text{r}$ and computed $C_\text{r}$ and $L_\text{r}$ for a random graph with a preserved degree distribution and an identical average number of edges per node \cite{Maslov2002}. 

In order to investigate synchronizability during seizures, we analyzed -- again in a time-resolved manner -- the spectrum of the Laplacian matrix $\bL$. As a measure for the stability of the globally synchronized state \cite{Nishikawa2003,Atay2006,Boccaletti2006a} of a connected graph we consider the eigenratio $S=\lambda_\text{max}/\lambda_\text{min}$, where $\lambda_\text{min}$ denotes the smallest non-vanishing eigenvalue and $\lambda_\text{max}$ the maximum eigenvalue of $\bL$. A network is said to be less synchronizable for larger values of $S$ and better synchronizable for smaller values of $S$, the latter indicating a more stable globally synchronized state \cite{Barahona2002,Nishikawa2003}.

\begin{figure}
  \includegraphics[width = 6cm]{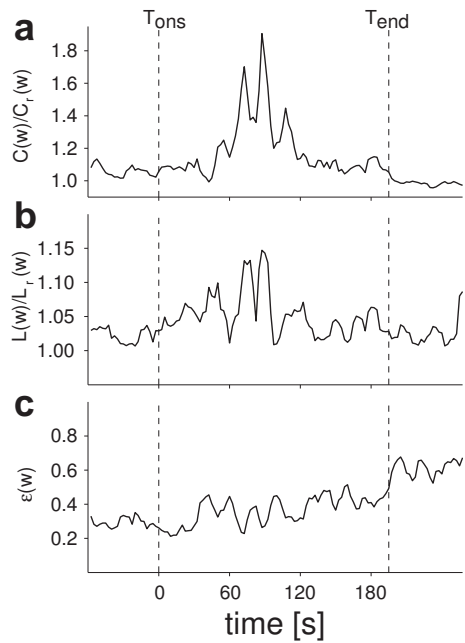}
  \caption{Evolving network properties during a seizure. Vertical broken lines indicate times of onset $T_\text{ons}$ and end $T_\text{end}$ of the seizure. 
  (\textbf{a}) Exemplary time course of the ratio $C(w)/C_\text{r}(w)$. The cluster coefficient $C(w)$ of the epileptic brain network is already slightly
  larger than the corresponding value of a random network $C_\text{r}(w)$ before seizure onset and attains a maximum deviation approximately in the
  middle of the seizure. Already prior to seizure end the cluster coefficient attains values indicative of a random network, which extends into the
  post-seizure period. A similar, though less pronounced temporal evolution can be observed for the average shortest path length (\textbf{b}). Edge density 
  $\epsilon(w)$ fluctuates around 0.3 during the first half of the seizure and slowly increases during the second half, which extends into the 		
  post-seizure period (\textbf{c}). Profiles are smoothed using a 4-point moving average.}
  \label{fig:figure2}
\end{figure} 

\section{Results}
\label{results}
In Fig. \ref{fig:figure2} we show typical time courses of $C(w)/C_\text{r}(w)$ and $L(w)/L_\text{r}(w)$. Given our thresholding, which can yield a different number of links for every time window, we show, in addition, the time course of $\epsilon(w)$. Already prior to the electrographic seizure onset (see Ref. \cite{Schindler2007a} for a fully automated detection of electrographic seizure onset and seizure end) both average shortest path length and cluster coefficient indicate a slight deviation from a random network, while edge density of the seizure network remains almost constant. Approximately in the middle of the seizure this deviation is most pronounced and indicates a movement toward a more regular functional topology. Interestingly, already prior to the electrographic seizure end we observe a movement away from the more regular functional topology, which extends into the post-seizure period.  Edge density slowly increases during the second half of the seizure and reaches an average value during the post-seizure period that is almost twofold the average value of the pre-seizure period. We note that this temporal evolution is also reflected in the dynamics of maximum degree $d_\text{max}(w)$. The difference between maximum and average degree ($\Delta d(w)= d_\text{max}(w)-d_\text{avg}(w)$) remains almost constant at one quarter of the number of nodes while the minimum degree $d_\text{min}(w)$ is 1 for all windows $w$ (data not shown). The variability of the concave-like temporal evolution of network characteristics $C$ and $L$ for different seizures from the same patient was low, and moreover, was a consistent finding for all investigated seizures independently of the anatomical location of their onset (cf. Figs. \ref{fig:figure4}a and b).

\begin{figure}
  \includegraphics[width = 6cm]{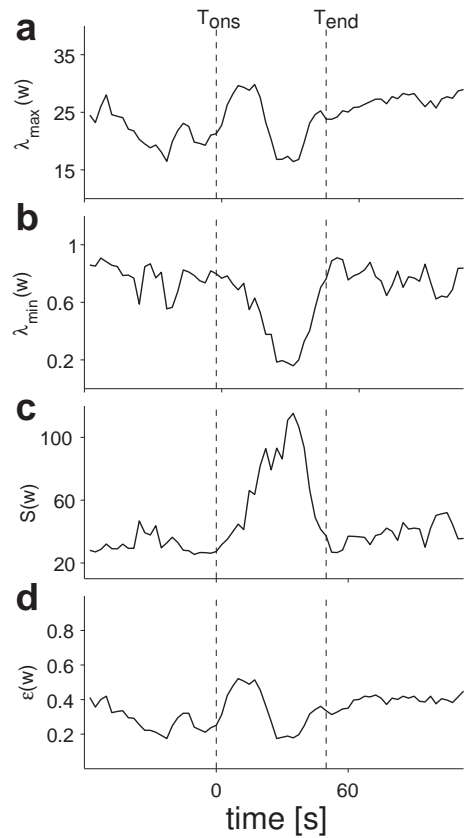}
  \caption{Evolving synchronizability during a seizure. Exemplary time courses of the largest eigenvalue $\lambda_\text{max}(w)$ (\textbf{a}), the smallest non-vanishing eigenvalue $\lambda_\text{min}(w)$ (\textbf{b}), the eigenratio $S(w)$ (\textbf{c}), and of edge density $\epsilon(w)$ (\textbf{d}). $S(w)$ mainly follows the dynamics of $\lambda_\text{min}(w)$ and shows a concave-like temporal evolution similar to the ones observed for the statistical network characteristics (cf. Fig. \ref{fig:figure2}). The dynamics of $\lambda_\text{max}(w)$ is largely dominated by $\epsilon(w)$. Vertical broken lines indicate times of onset $T_\text{ons}$ and end $T_\text{end}$ of the seizure. Profiles are smoothed using a 4-point moving average.}
  \label{fig:figure3}
\end{figure}

\begin{figure}
  \includegraphics[width = 6cm]{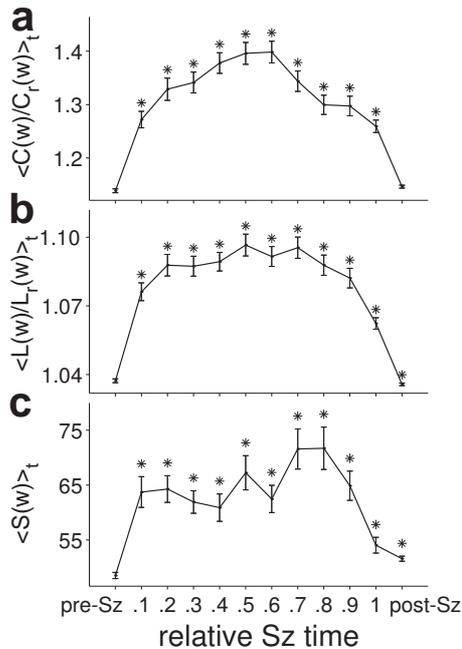}
  \caption{(\textbf{a}) Means of $C/C_\text{r}$ averaged separately for pre-seizure, discretized seizure, and post-seizure time periods of 100 epileptic seizures. Seizures (mean duration: $110 \pm 60$ s) were partitioned into 10 equidistant bins. Error bars indicate standard error of the mean. (\textbf{b}) Same as \textbf{a} but for the $L/L_\text{r}$. (\textbf{c}) Same as \textbf{a} but for the eigenratio $S$. Stars denote significant changes relative to the pre-seizure time periods ($p<0.01$; Bonferroni corrected pairwise Wilcoxon rank sum tests for equal medians). Lines connecting the mean values are for eye guidance only.}
  \label{fig:figure4}
\end{figure}

Typical time courses for $\lambda_\text{max}(w)$, $\lambda_\text{min}(w)$, and $S(w)$ during a seizure are shown in Fig. \ref{fig:figure3}. Again we observed a concave-like temporal evolution, with highest values of $S$ (i.e., lowest synchronizability) in the middle of the seizure, followed by a decline (i.e., an increasing synchronizability) already prior to the electrographic seizure end. Although this behavior varied from patient to patient, it was a consistent finding for all seizures (cf. Fig. \ref{fig:figure4}c). A comparison of Fig. \ref{fig:figure3}b with Fig. \ref{fig:figure3}c shows that $S(w)$ is largely dominated by the dynamics of the smallest non-vanishing eigenvalue $\lambda_\text{min}(w)$, and its decrease in the middle of the seizure may indicate a reorganization of the network into local sub-structures \cite{Atay2006,Huang2006}. In this case, sparsely occurring links between local sub-structures can significantly affect $\lambda_\text{min}$ \cite{Huang2008}. The relative change of largest eigenvalue $\lambda_\text{max}(w)$ during the seizure is less pronounced as compared to that of $\lambda_\text{min}(w)$ and resembles the time course of edge density $\epsilon(w)$ (cf. Fig. \ref{fig:figure3}d). This similarity can be expected, at least to some extent, since edge density constitutes a lower bound for the largest eigenvalue, $\lambda_\text{max} \geq n\epsilon$ (cf. \cite{Atay2005}). We could, however, not observe such a clear cut influence of $\epsilon$ on $\lambda_\text{min}$ and hence on $S$. As regards the degree distribution, we again observed $\Delta d(w)\approx 0.25n$ and $d_\text{min}(w)=1$ for all windows and a temporal evolution of $d_\text{max}(w)$ quite similar to the dynamics of $\epsilon(w)$ (data not shown) indicating that the degree distribution of the seizure network does not appear to determine its synchronizability (cf. \cite{Atay2006b}). 

Fig. \ref{fig:figure4} summarizes the dynamics of functional network properties and synchronizability for all one-hundred focal onset seizures. Irrespective of the anatomical location of seizure origin, both the normalized cluster coefficient and the normalized average shortest path length rapidly increased during the first half of the seizures then gradually decreased again. Interestingly, this temporal evolution was more pronounced for the normalized cluster coefficient than for the normalized average shortest path length. This indicates a relative shift toward a less random functional topology of the seizure state. Seizures are usually associated with massively synchronized brain activity \cite{Schindler2007a,Schindler2007b}, and the significantly decreased synchronizability of the underlying functional topology may catalyze the emergence of a globally synchronized state of the epileptic brain. Once such a state has been established, synchronizability increases again, as observed in our data.

\begin{figure}
	\includegraphics[width = 7.8cm]{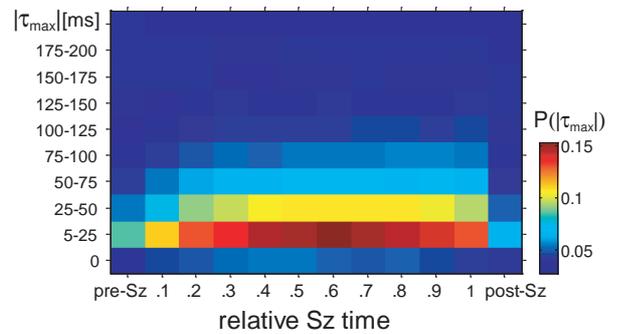}
  \caption{(Color online) Normalized frequency distribution $P\left(\left|\tau_\text{max}\right|\right)$ of absolute time lag of maximum
  cross-correlation for all seizures. Given the duration of an analysis window 
  $\left|\tau_\text{max}\right|$ is confined to the interval [0,2499] ms. Note the finer resolution for smaller time lags. 
  We here show the range $\left|\tau_\text{max}\right| \in [0, 225]$ ms only, since $P\left(\left|\tau_\text{max}\right|\right) \rightarrow 0$ for larger
  $\left|\tau_\text{max}\right|$.}
  \label{fig:figure5}
\end{figure} 

If the observed changes of functional topology were simply a consequence of enhanced volume conduction during the seizures, i.e., due to direct propagation of distant sources to remote sensors, the covariance of the EEG signals would be expected to occur with zero time lag. In order to exclude the effect of a linear superposition of sources leading to a more regular-looking graph, we estimated the normalized frequency distribution of absolute time lag of maximal correlation $P\left(\left|\tau_\text{max}\right|\right)$ (cf. Eq. (\ref{eq:rho})) for all seizures, partitioned into 10 equidistant time bins (cf. Fig. \ref{fig:figure4}). We observed $P\left(\left|\tau_\text{max}\right|\right)$ to peak in the range of 5--50ms (see Fig. \ref{fig:figure5}), which indicates that the observed changes of functional topology are not due to passive electromagnetic field effects in the extracellular space, but rather due to propagation of electrical activity along anatomical pathways. 

\section{Conclusion}
\label{conclusion}
We have presented findings obtained from a time-resolved analysis of statistical and spectral properties of functionally defined networks underlying human epileptic seizures. Despite the many influencing factors (number of nodes, non-uniform arrangement of sensors, focus on particular brain regions, choosing a threshold for the extraction of functional networks, etc.) that impede an interpretation of graph-theoretical measures in a strict sense when analyzing field data, our results indicate that seizure dynamics -- irrespective of the anatomical onset location -- can be characterized by a relative transient shift toward a more regular and then back toward a more random functional topology. This is consistent with recent observations reported in Ref. \cite{Ponten2007}, where a small number of seizures that originated from a circumscribed brain region have been analyzed.

We here observed that the changing functional network topology during seizures was accompanied by an initially decreased stability of the globally synchronized state, which increased already prior to seizure end. In a previous study \cite{Schindler2007a} we analyzed the same data set using multivariate time series analysis techniques from random matrix theory and observed that -- surprisingly -- global neuronal synchronization (derived from the eigenvalue spectrum of the zero-lag correlation matrix) significantly increased during the second half of the seizures, and before seizures stopped. Our present findings indicate that such a global increase of neuronal synchronization prior to seizure end may be promoted by the underlying functional topology of brain dynamics. This corroborates the hypothesis \cite{Topolnik2003,Schiff2005,Schindler2007a,Schindler2007b} that increasing synchronization of neuronal activity may be considered as an emergent self-regulatory mechanism for seizure termination. Thus, our result can provide clues as to how to control seizure network, e.g. via pinning \cite{Grigoriev1997,Sorrentino2007}. 

While the aforementioned interpretation would indicate that the transient evolution in graph properties is an active process of the brain to abort a seizure, our findings could also be understood as a passive consequence of the seizure itself. The extremely intense firing of neurons during a seizure might lead to a saturation of the capacity of neurons to fire, particularly in brain areas with a high number of functional links. If such hubs are saturated and become silent, then the connections between local sub-structures are affected to a larger extent, which could lead to a segregation of the global network and to a decrease of global synchronizability. Thus, a reliable identification of local sub-structures or hubs could improve understanding of mechanisms underlying the generation, maintenance, propagation, and termination of epileptic seizures.

At present, our findings are restricted to interdependences that can be assessed by the cross-correlation function. Future studies will clarify whether additional information can be gained from analyses invoking other time series analysis techniques, including nonlinear ones, as well as techniques that take into account the direction of interactions. Nevertheless, disentangling the interplay between connection structure and dynamics of the complex network human brain may advance our understanding of epileptic processes.

\begin{acknowledgments}
We thank Philip H. Goodman and Christof Cebulla for helpful comments. K. S. was supported by a scholarship of the SSMBS (Schweizerische Stiftung f\"ur Medizinisch-Biologische Stipendien) donated by Roche. S. B. was supported by the German National Academic Foundation (Studienstiftung). M.-T. H., C. E. E., and K. L. acknowledge support from the Deutsche Forschungsgemeinschaft (Grant Nos. SFB-TR3 sub-project A2 and LE660/4-1).
\end{acknowledgments}

\end{document}